\documentclass{PoS}

\usepackage{amsmath}
\usepackage{amsfonts}
\usepackage{amssymb}
\usepackage{dsfont}

\title{On observable particles in theories with a Brout-Englert-Higgs effect}

\ShortTitle{On observable particles in theories with a Brout-Englert-Higgs effect}

\author{\speaker{Pascal T\"orek}
        \thanks{Partly supported by the FWF doctoral school W1203-N16.
The computational results presented have been obtained using the Vienna Scientific Cluster (VSC), 
the HPC center at the University of Graz and the Graz University of Technology. We are grateful to Ren\'e Sondenheimer for a critical reading.}
        \\
        Institute of Physics, NAWI Graz, University of Graz, Universit\"atsplatz 5, 8010 Graz, Austria\\
        E-mail: \email{pascal.toerek@uni-graz.at}}

\author{Axel Maas\\
        Institute of Physics, NAWI Graz, University of Graz, Universit\"atsplatz 5, 8010 Graz, Austria\\
        E-mail: \email{axel.maas@uni-graz.at}}

\abstract{Even at weak coupling the physical, observable spectrum of gauge theories with a 
Brout-Englert-Higgs effect can deviate from the elementary one of perturbation theory. 
This can be analytically described and treated using the Fr\"ohlich-Morchio-Strocchi mechanism. 
We confirm this by lattice simulation for an $\mathrm{SU}(3)$ gauge theory with a fundamental scalar, 
a toy model for grand unification. We also show that this has experimentally observable consequence, 
e.g., in scattering cross-sections of lepton collisions in this toy model.
}

\FullConference{An Alpine LHC Physics Summit (ALPS2018)\\
		15-20 April, 2018\\
		Obergurgl, Austria}

\begin{document}

\section{Introduction}

The physical spectrum of gauge theories need to be manifestly gauge invariant. In the standard model 
this would imply that also the W/Z bosons, the Higgs, and the fermions, all gauge-variant under the weak 
gauge group, are unphysical \cite{'tHooft:1979bj,Frohlich:1980gj,Frohlich:1981yi}. Nevertheless, 
a standard perturbative treatment within a fixed gauge, treating the elementary fields as if they were 
observable, describes experimental results remarkably well \cite{pdg}. 

This apparent contradiction is resolved by the mechanism described by Fr\"ohlich, Morchio, and Strocchi (FMS) 
\cite{Frohlich:1980gj,Frohlich:1981yi}: Under certain conditions, fulfilled by the standard model, 
the properties of the physical states are in a one-to-one correspondence to those of the gauge-variant 
elementary particles. This mechanism has been confirmed in lattice simulations of the weak-Higgs sector of the
standard model \cite{Maas:2013aia,Maas:2012tj}, and exploratively in the full Georgi-Glashow-Weinberg 
model \cite{Shrock:1985un,Shrock:1985ur}. 

However, the one-to-one correspondence relies on the special structure of the standard model, where the gauge 
group and the global (custodial) symmetry group are the same, $\mathrm{SU}(2)$. Therefore, standard perturbation 
theory alone may or may not be sufficient to describe beyond the standard model (BSM) physics\cite{Maas:2015gma}. 
Fortunately, gauge-invariant perturbation theory, constructed upon the FMS mechanism 
\cite{Torek:2016ede,Maas:2017pcw,Maas:2017wzi}, promises to be still able to give analytical access to the 
phenomenology, even if standard perturbation theory fails.

Here we test gauge-invariant perturbation theory and the FMS mechanism for BSM-like structures as they appear, 
e.g., in grand-unified theories (GUTs). We consider an $\mathrm{SU}(3)$ gauge theory  with a single fundamental 
scalar. We give predictions in Section \ref{s:pred} which are successfully tested against lattice 
calculations \cite{Maas:2016ngo,Maas:2018xxu} in Section \ref{s:lattice}. In Section \ref{s:scattering}, 
we continue to discuss how experimentally observable consequences arise in a toy-version of a lepton collider 
in this theory.

\section{SU(3) gauge theory with a fundamental scalar}\label{s:pred}

We study an $\mathrm{SU}(3)$ gauge theory with a fundamental scalar and action
\begin{align}
S &= \int \mathrm{d}^4 x\Big[-\frac{1}{4}F_{\mu\nu}^aF^{a\,\mu\nu}
+\big(D_\mu\phi\big)^\dagger\big(D^\mu\phi\big)-V\big(\phi^\dagger\phi\big)\Big]\;.
\label{eq:contaction}
\end{align}
Herein the gauge fields $A_\mu$ with field-strength tensor $F_{\mu\nu}$ couple to the scalar field 
$\phi$ through the covariant derivative $D_\mu$. Since the potential $V$ depends only on the 
gauge-invariant combination $\phi^\dagger\phi$, the theory exhibits a global $\mathrm{U}(1)$ 
symmetry acting only on the scalar. This is the custodial symmetry of the theory \cite{Maas:2017pcw}. 

\subsection{Gauge-variant spectrum}\label{ssec:gv}

We first concentrate on a tree-level analysis to investigate the mass spectrum of the elementary 
fields in a fixed gauge with a non-vanishing vev for the scalar field. This will provide the predictions 
of standard perturbation theory.

Therefore, we split the scalar field into its vev $v$, 
which minimizes the potential $V$, and a fluctuation part $\varphi$ around the vev,  i.e.,
\begin{align}
\phi(x) &= \frac{v}{\sqrt{2}}n+\varphi(x)\;,
\label{eq:split}
\end{align}
where $n$ is a unit vector in gauge space pointing in the direction of the vev. Without loss of 
generality we set $n=\delta_{i,3}$. This  can be achieved by, e.g., using the 't Hooft gauge 
condition \cite{Bohm:2001yx}. The Higgs boson and the Goldstone bosons can be described 
in a gauge-covariant (but not gauge-invariant) manner without specifying $n$ by 
$h=\sqrt{2}\mathrm{Re}[n^\dagger\phi]$ and $\hat{\varphi}=\phi-\mathrm{Re}[n^\dagger\phi]n$. 

Inserting the split \eqref{eq:split} into the kinetic term of scalar field in \eqref{eq:contaction} 
yields the mass matrix for the gauge bosons
\begin{align}
\big(M_\mathrm{A}^2\big)^{ab} &= \frac{g^2v^2}{4}\mathrm{diag}
\Big(0,0,0,1,1,1,1,\frac{4}{3}\Big)^{ab}\;.
\end{align}
Therefore, we obtain $3$ massless gauge bosons, $4$ degenerate massive gauge bosons with mass 
$m_\mathrm{A}=gv/2$, and one gauge boson with mass 
$M_\mathrm{A}=\sqrt{4/3}~m_\mathrm{A}>m_\mathrm{A}$. Additionally, from the quadratic terms in 
$\varphi$ of the potential one obtains a mass $m_h=\lambda v$ for the elementary Higgs field, where 
$\lambda$ is the usual four-Higgs coupling. 

The situation is now that which, in an abuse of language, is usually called 'spontaneously broken'. 
The breaking pattern is $\mathrm{SU}(3)\to\mathrm{SU}(2)$. 

\subsection{Gauge-invariant spectrum}\label{ssec:gi}

The gauge-invariant, and thus experimentally observable, spectrum of the theory, can be obtained using 
gauge-invariant perturbation theory using the FMS mechanism, see \cite{Torek:2016ede,Maas:2017pcw} for 
a description of the procedure.

The states are classified by the quantum numbers $J^{PC}_{\mathrm{U}(1)}$, 
where $J$ is the total angular momentum, $P$ the parity, $C$ the charge parity, and the lower 
index corresponds to the global $\mathrm{U}(1)$ charge. Therefore, the states are either singlets 
or non-singlets with respect to the custodial group. 

First, let us concentrate on the singlet states \cite{Maas:2016ngo}: A gauge-invariant 
operator describing a scalar, positive (charge-) parity boson is 
$O_{0^{++}_0}(x)=(\phi^\dagger\phi)(x)$. 
We apply the FMS mechanism and expand the subsequent correlation functions to leading order, giving 
\cite{Torek:2016ede}
\begin{align}
\big\langle O_{0^{++}_0}(x)O_{0^{++}_0}(y)^\dagger\big\rangle &= 
\frac{v^4}{4} + v^2\big\langle h(x)h(y)\big\rangle_{\mathrm{tl}} + 
\langle h(x)h(y)\big\rangle_{\mathrm{tl}}^2 +\cdots\;,
\label{eq:fms0pp0}
\end{align}
where 'tl' means 'tree-level'. The second term on the r.h.s. of Equation \eqref{eq:fms0pp0} 
describes the propagation of a single elementary Higgs boson and the third term two non-interacting 
Higgs bosons. Comparing poles on both sides predicts the mass of the bound state $O_{0^{++}_0}$, 
i.e., the observable particle. This scalar boson should therefore have a mass equal to the mass of the 
elementary Higgs. The next state in this channel is a scattering state with twice the mass of the 
elementary Higgs. 

Now, we focus on a singlet vector operator 
$O_{1^{--}_0}^\mu(x)=\mathrm{i}(\phi^\dagger D^\mu\phi)(x)$. Using again the FMS prescription 
yields 
\begin{align}
\big\langle O_{1^{--}_0}^\mu(x)O_{1^{--}_0,\mu}(y)^\dagger\big\rangle &=\label{vec}
\frac{v^4g^2}{12}\big\langle A^{8\,\mu}(x)A_{\mu}^{8}(y)\big\rangle_{\mathrm{tl}}+\cdots\;.
\end{align}
The poles of the r.h.s. is at the mass $M_\mathrm{A}$ of the heaviest gauge boson. Thus a single 
massive state is predicted in this channel, which contradicts the perturbative expectation in 
Section \ref{ssec:gv}.

Next, we study gauge-invariant states with open custodial quantum numbers \cite{Maas:2017xzh}. 
Since the corresponding charge is conserved, the lightest such state is absolutely stable. Note, 
that there is no elementary field in the theory with these quantum numbers and thus is not predicted 
by standard perturbation theory. A scalar operator and vector operator with open custodial charge 
are given by
\begin{align}
O_{0^{++}_1} &= \epsilon_{ijk}\,\phi_i\,\big(D_\mu\phi\big)_j\,\big(D_\mu D^2\phi\big)_k 
\quad\text{and}\quad 
O_{1^{--}_1}^\mu = \epsilon_{ijk}\,\phi_i\,\big(D_\mu\phi\big)_j\,\big(D^2\phi\big)_k\;.\label{dm}
\end{align} 
Applying the FMS mechanism and employing a straightforward, but tedious, tree-level analysis to these bound state 
correlators, see \cite{Maas:2017xzh}, gives for both a ground state mass of $2m_\mathrm{A}$.

\section{Results from lattice simulations}\label{s:lattice}

The predictions above are dramatically different from the usual perturbative predictions. They thus require 
close scrutiny. In absence of experiment, and the relevance of bound states, a possibility are genuine 
non-perturbative methods. Our tool of choice here are lattice simulations, though any other 
non-perturbative method could (and should) be used as well.

We discretize the action \eqref{eq:contaction} on a $4$-dimensional, Euclidean lattice with lattice 
size $L$ and lattice spacing $a$, in a standard way, see 
\cite{Montvay:1994cy}. We then perform Monte Carlo simulations for 
different lattice sizes and lattice parameters, see \cite{Maas:2016ngo,Maas:2018xxu} for details of the simulations.

\begin{figure}[tbh!]
\begin{center}
\includegraphics[width=0.5\textwidth]{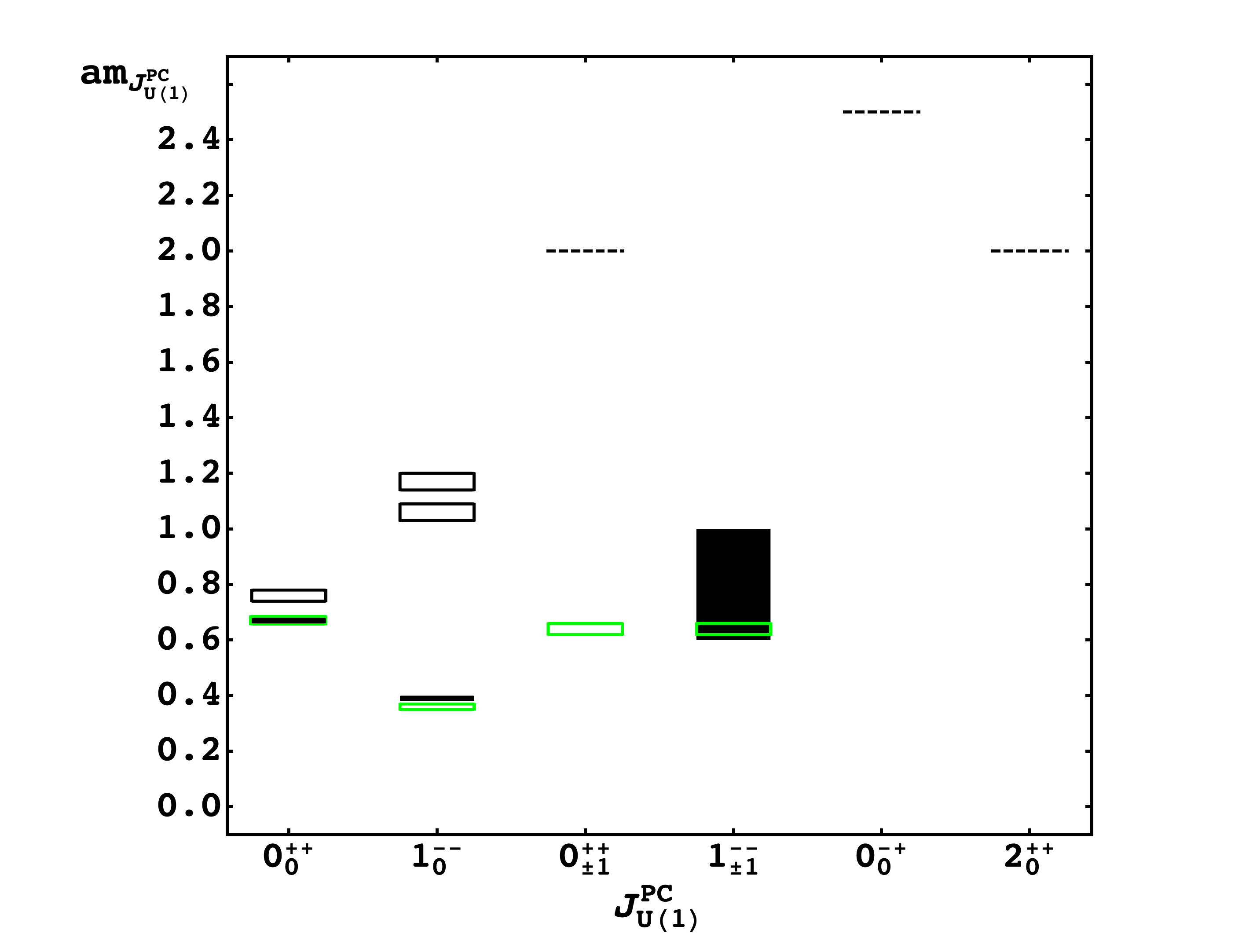}\includegraphics[width=0.5\textwidth]{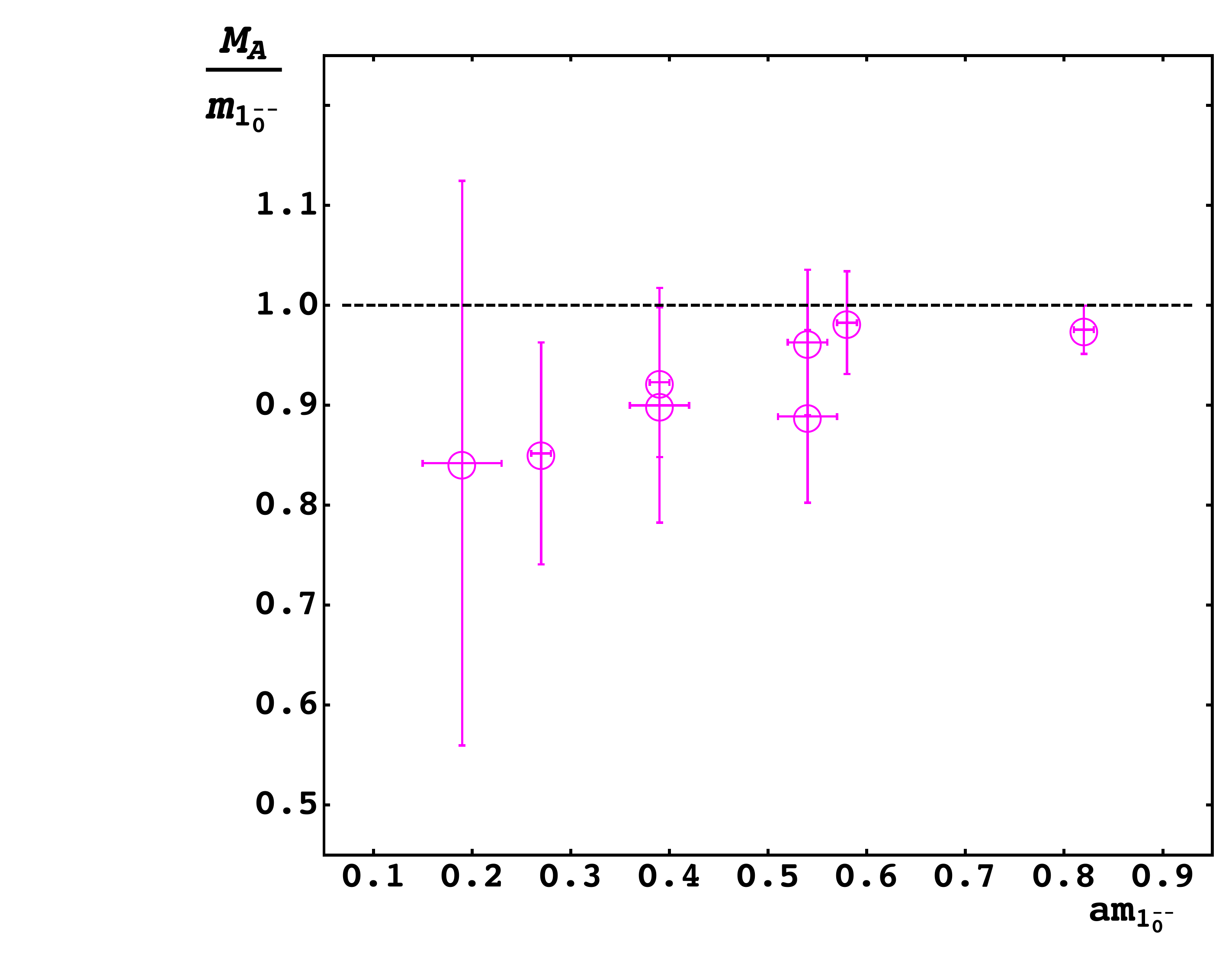}
\caption{Left panel: Physical (gauge-invariant) spectrum of the theory (black boxes and dashed lines) 
compared to the predictions of the FMS mechanism to leading order (green boxes). Right panel: Ratio of the 
mass of the physical vector singlet to the predicted mass as a function of the mass of the physical vector 
singlet in lattice units. Both results from \cite{Maas:2018xxu}.
\label{fig:spectrum}}
\end{center}
\end{figure}

We determined both the gauge-dependent correlators of Section \ref{ssec:gv} in a fixed gauge, as well as the 
correlators of the composite states in Section \ref{ssec:gi}.

From the gauge-variant correlators we extracted $m_\mathrm{A}$ and $M_\mathrm{A}$. We find the ratio of these 
masses to be in good agreement with leading-order perturbation theory \cite{Maas:2016ngo,Maas:2018xxu}. This is 
in agreement with the fact that the running coupling was also found to be small \cite{Maas:2018xxu}. This 
provided the pole-position on the right-hand side of \eqref{vec}, as well as the predictions for the masses 
of the states (\ref{dm}). Likewise, we extracted from the gauge-invariant correlators, using a 
variational analysis, the low-lying spectrum in several quantum number channels \cite{Maas:2016ngo,Maas:2018xxu}. 

On the left panel in Figure \ref{fig:spectrum} we show for a sample set of parameters of the 
action \eqref{eq:contaction} the physical spectrum of the theory for different quantum number channels 
\cite{Maas:2018xxu}. The full black boxes are the extracted ground states, the empty boxes are the elastic 
thresholds for the scalar and vector singlet channels (the vertical extent of these boxes shows the statistical 
error), and the dashed lines are estimated\footnote{No reliable result has been obtained in these cases, since 
the statistics is too low in these noisy channels.} ground state masses of the $0^{++}_{\pm1}$, $0^{-+}_0$, 
and $2^{++}_0$ channels. The results are compared to the predictions of the FMS mechanism and gauge-invariant 
perturbation theory of Section \ref{ssec:gi} (green boxes). The latter used the gauge-dependent results as input.

The overall agreement shows that the spectrum is, even to leading order, well predicted\footnote{The scalar 
non-singlet is too noisy and more statistics is needed to make definite statements.}. Given that there are 
still substantial lattice artifacts present \cite{Maas:2018xxu}, and that the calculations in 
Section \ref{ssec:gi} and \cite{Maas:2017xzh} are just tree-level calculations, the agreement is remarkably good. 
This good agreement was confirmed also for other sets of parameters \cite{Maas:2018xxu}, which is shown on the 
right panel in Figure \ref{fig:spectrum} for the case of the vector singlet channel.

\section{Scattering processes}\label{s:scattering}

The results in Section \ref{s:lattice} support the FMS mechanism and show that gauge-invariant perturbation theory 
\cite{Maas:2017wzi} is able to predict the spectrum analytically. However, in actual experiments gauge bosons 
and the Higgs are never observed as final states, but only as intermediate states. It is 
therefore interesting to study, whether the qualitative differences observed in the spectrum carries through 
to an experiment, which uses handleable (fermionic) matter as initial and final states.

Scattering processes of such fermions can also be described using gauge-invariant perturbation theory 
\cite{Maas:2012tj,Maas:2017wzi,Egger:2017tkd}. E.g., this can be used to explain why the bound-state nature of 
the physical states in the standard model has escaped detection so far \cite{Maas:2017wzi,Egger:2017tkd}. But 
here we are interested in how deviations could arise.

We therefore couple a fermion $\psi$ in the fundamental representation to the theory \eqref{eq:contaction}. For 
the moment, it will not matter whether they are chirally or vectorially coupled, nor will it be relevant if they 
are interacting with the Higgs through a Yukawa interaction.

Following the rules of the FMS mechanism \cite{Frohlich:1981yi,Maas:2017wzi,Egger:2017tkd} a physical fermion-number one 
state is obtained by the fermionic bound state $\Psi=\phi^\dagger\psi$. This state carries, besides the fermion 
number, a U(1) custodial charge, which takes over the role of the would-be flavor charge. Hence, there is only 
a single physical fermion state. To leading order, its propagator expands like
\begin{align}
\big\langle(\phi^\dagger\psi)^\dagger(x)\gamma_0(\phi^\dagger\psi)(y)\big\rangle &= 
v^2\big\langle\bar{\psi}_3(x)\psi_3(y)\big\rangle_\mathrm{tl}+\cdots\;.
\end{align}
Thus, the ground state in this channel has the mass of the perturbative fermion state in the broken subsector.
While the appearance of the state is consistent with perturbation theory, it should be noted that no analogue 
state of the other two fermions $\psi_{1,2}$ appear, in contrast to perturbation theory. They only appear as quantum corrections. Thus, there is again a qualitatively difference 
in the fermion-number one sector\footnote{There are also three-fermion states, similar to nucleons. They carry a three-times larger fermion number and thus belong to a different charge sector \cite{Egger:2017tkd}, 
and will therefore be ignored here.}.

A physical scattering process of two such fermions to two such fermions will now be \cite{Egger:2017tkd}
\begin{align}
\big\langle\bar{\Psi}(p_1)\bar{\Psi}(p_2)\Psi(q_1)\Psi(q_2)\big\rangle &=
v^4\big\langle\bar{\psi}_3(p_1)\bar{\psi}_3(p_2)\psi_3(q_1)\psi_3(q_2)\big\rangle_\mathrm{tl}+\cdots\;.
\end{align}
The color structure of the interaction is given by the Gell-Mann matrices. To leading order the expression 
involves $\lambda_{33}^aD_{\mu\nu}^{ab}\lambda_{33}^b$, where $\lambda^a$ are the Gell-Mann matrices and 
$D_{\mu\nu}^{ab}$ is the gauge boson propagator \cite{Bohm:2001yx}. But only $\lambda^8$ has a non-vanishing 
$33$ component, and thus only the propagator $D^{88}$ contributes. Hence, in the scattering cross-section only 
poles at $M_A$ arises. But this is exactly the mass of the physical vector state! Thus, even in the cross-section measured in an experiment at leading order only the physical state shows up as a resonance. This
supports that only the gauge-invariant states are observable. Of course, investigations beyond leading order are needed to confirm this.

\section{Conclusions}

In the case presented here and in \cite{Maas:2018xxu}, as well as in all cases investigated so far 
\cite{Maas:2017wzi}, gauge-invariant perturbation theory and the FMS mechanism has provided correct predictions. 
This explains both the success of standard perturbation theory, as well as its failures in situations as those 
presented here. Investigations like the one in Section \ref{s:scattering} and \cite{Maas:2017wzi,Egger:2017tkd} 
also show that this transfers through to experimentally accessible quantities, like cross-sections. Thus, a 
treatment using gauge-invariant perturbation theory seems to be the suitable approach to theories with a BEH 
effect.

\bibliographystyle{bibstyle}
\bibliography{bib}

\end{document}